\documentclass[twocolumn,prl,superscriptaddress,unsortedaddress,showpacs,preprintnumbers,amsmath,amssymb]{revtex4}
\usepackage[dvips]{graphicx}
\usepackage{dcolumn}
\usepackage{amsmath}
\usepackage{times}

\begin{document}


\title{  $e^+ + H$ direct annihilation above the positronium
formation threshold }

\author{C-.Y. Hu, D. Caballero, A. Forrester and Z. Papp}

\affiliation{
 Department of Physics and Astronomy, 
California State University, Long Beach, California 90840 }

\date{\today}

\begin{abstract}
A long-standing problem with the solution of the Schr\"odinger equation has been
its inability to account for the electron-positron annihilation in positron
hydrogen scattering above the positronium formation threshold. This letter shows
that this problem has been resolved by the use of the modified Faddeev
equations. A number of $e^- e^+$ annihilation
cross sections in the energy gap between $Ps(1s)$ and $H(n=2)$ thresholds are
reported for both the positron plus hydrogen incoming channel as well as the
proton plus positronium incoming channel. However the indirect 
annihilation cross
sections after formation of the positronium 
themselves are well known, they will not be included in
this report.
\end{abstract}


\maketitle

When the possibility of $e^+ e^-$ annihilation in flight is neglected the 
$e^+ +H$ collision in the energy region between $Ps(1s)$ and $H(n=2)$ thresholds
can have only two open channels. They are: the elastic channel and the
rearrangement channel, where the $e^+$ capture the electron from $H$ and form a
positronium in $Ps(1s)$. Under this assumption annihilation can occur only after
the formation of the positronium. The lifetime of the $Ps(1s)$ is known to
depend on the total spin $S$. 
The singlet para-$Ps$, $^1S_0$, decays into $2\gamma$ rays with a lifetime
$\tau \approx 0.13 ns$. The triplet ortho-$Ps$, $^3S_1$, decay into $3\gamma$ 
rays with $\tau \approx 140 ns$ \cite{rich}. These well known annihilation will not
be included in this report. We calculate the $e^+ e^-$ direct 
annihilation 
using the standard relation \cite{fraser}
\begin{equation}
\sigma_{an}=\pi r_0^2 (c/v) Z_{\mbox{eff}},
\label{eqn1}
\end{equation}
$r_0=e^2/m_e c^2$, $Z_{\mbox{eff}}$ is the annihilation coefficient defined by
\begin{equation}
Z_{\mbox{eff}} = \langle \Psi | \delta(\vec{r}) \Psi \rangle,
\label{eqn2}
\end{equation}
$\vec{r}$ is the distance between $e^+ e^-$ in the three-body wave function
$\Psi$. The calculation of $e^+ e^-$ in flight thus compensate for the 
shortcoming of the quantum three-body calculation where the 
annihilation channel is closed.

The modified Faddeev equation  \cite{merkuriev,hu} explicitly separates
the three-body wave function into the direct channel and the rearrangement
channel
\begin{equation}
\Psi = \Psi_1(\vec{x}_1,\vec{y}_1) +  \Psi_2(\vec{x}_2,\vec{y}_2).
\end{equation}
Here $\vec{x}_\alpha,\vec{y}_\alpha$, $\alpha=1,2$ are the corresponding 
mass-scaled Jacobi vectors \cite{hu} defined by
\begin{eqnarray}
\vec{x}_\alpha & =& \tau_\alpha(\vec{r}_\beta-\vec{r}_\gamma) \\ 
\vec{y}_\alpha & =& \mu_\alpha\left(\vec{r}_\alpha -
\frac{m_\beta\vec{r}_\beta+m\gamma\vec{r}_\gamma}{m_\beta+m_\gamma}\right), 
\nonumber
\end{eqnarray}
where $(\alpha\beta\gamma)$ are cyclic permutations of $(123)$, $m_\alpha$ and
$\vec{r}_\alpha$ are the particle mass and position vectors,
\begin{equation}
\tau_\alpha=\sqrt{\frac{2 m_\beta m_\gamma}{m_\beta+m_\gamma}},\ \ \ \ \ 
\mu_\alpha=\sqrt{2 m_\alpha \left( 1-\frac{m_\alpha}{M} \right)},
\end{equation}
and $M=m_\alpha+m_\beta+m_\gamma$.
 
\begin{figure}
\resizebox{8cm}{!}{
\includegraphics{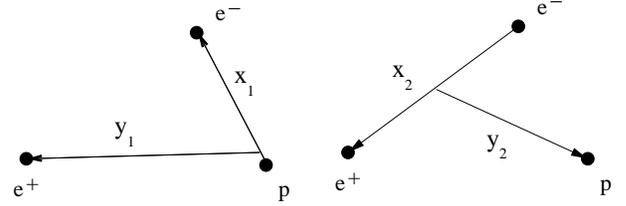}}
\caption{Jacobi vectors in the $e^+ -H$ system.}
\label{jacobi}
\end{figure}

The Jacobi vectors corresponding to different channels are related by the
orthogonal transformation
\begin{equation}
\left(\begin{array}{c} \vec{x}_\alpha
 \\ \vec{y}_\alpha \end{array} \right) =
\left(\begin{array}{cc} 
C_{\beta \alpha}  & S_{\beta \alpha}  \\ 
 - S_{\beta \alpha} & C_{\beta \alpha} 
\end{array} \right) 
\left(\begin{array}{c} \vec{x}_\beta
 \\ \vec{y}_\beta \end{array} \right) 
 \label{eqn6}
\end{equation}
with $$C_{\beta \alpha} = - 
\left[ \frac{m_\beta m_\alpha}{(M-m_\beta)(M-m_\alpha)} \right]^{1/2}$$ and
$$S_{\beta \alpha} = (-)^{\beta-\alpha}
\mbox{sgn}(\alpha-\beta)(1-C^2_{\beta\alpha})^{1/2}.$$
 In this calculation we used the Jacobi vectors 
given in Fig.\ \ref{jacobi}. 

Using bipolar expansion we have 
\begin{equation}
\psi_\alpha ( {\bf x}_\alpha ,{\bf y}_\alpha)=
    \sum_{L=0}^{\infty}  \sum_{M=-L}^L 
\sum_{\vec{l}+\vec{\lambda}=\vec{L}} 
\frac{\psi_{\alpha_{l \lambda}}^L 
({x}_\alpha, {y}_\alpha) }{{x}_\alpha {y}_\alpha} 
Y_{l \lambda}^{L M} (\hat{x}_\alpha, \hat{y}_\alpha) 
\label{eqn7}
\end{equation}
where $\alpha=1,2$ and the bipolar basis is
$$Y_{l \lambda}^{L M} (\hat{x}_\alpha, \hat{y}_\alpha)=   
\sum_{m_l+m_\lambda=M} \langle l m_l \lambda m_\lambda | L M \rangle
Y_{l}^{m_l} (\hat{x}_\alpha )  Y_{\lambda}^{m_\lambda} (\hat{y}_\alpha ).
$$
Here $L$ is the total angular momentum of the three-body system and
$l$, $\lambda$ are the relative angular momenta corresponding to the Jacobi
vectors $\vec{x}_\alpha$ and $\vec{y}_\alpha$, respectively.
The modified Faddeev equations for angular momentum $L$ are given by
\begin{widetext}
\begin{equation}
\left[ H^{(\alpha)}_{l \lambda} + {V}_\alpha -E \right] 
\psi_{\alpha l \lambda}^L (x_\alpha,y_\alpha)
+ \sum_{\vec{\l}'+\vec{\lambda}'} W^{\alpha L}_{l\lambda,l'\lambda'}
\psi_{\alpha l' \lambda'}^L (x_\alpha,y_\alpha)
=
-V_\alpha^{(s)} 
\sum_{\vec{\l}'+\vec{\lambda}'} (h^{\alpha \beta  L}_{l\lambda,l'\lambda'}
\psi_{\beta l' \lambda'}^L) (x_\alpha,y_\alpha),
\label{eqn8}
\end{equation}
\end{widetext}
where
\begin{equation*}
H^{(\alpha)}_{l \lambda} =-\partial^2_{x_\alpha} -\partial^2_{y_\alpha}
+\frac{l(l+1)}{x_{\alpha}^2}  +\frac{\lambda(\lambda+1)}{y_{\alpha}^2},  
\end{equation*}
$V_\alpha$ is the Coulomb potential,
\begin{equation*}
W^{\alpha L}_{l\lambda,l'\lambda'}=
\langle Y_{l\lambda}^{LM} (\hat{x}_\alpha,\hat{y}_\alpha)| \bar{V}_\alpha |
Y_{l'\lambda'}^{LM} (\hat{x}_\alpha,\hat{y}_\alpha) \rangle,
\end{equation*}
\begin{equation*}
\bar{V}_{\alpha} =V_3(x_3)+V_\beta^{(l)}(x_\beta,y_\beta)
\end{equation*}
\begin{equation*}
V_{\alpha}^{(s)} =V_{\alpha} \zeta_\alpha (x_\alpha,y_\alpha)
\end{equation*}
\begin{equation*}
V_{\alpha}^{(l)} =V_{\alpha} (1- \zeta_\alpha (x_\alpha,y_\alpha)),
\end{equation*}
\begin{equation*}
\zeta (x,y))=2\left\{1+\exp\left[ \frac{(x/x_0)^\nu}{1+y/y_0}\right]\right\}^{-1},
\end{equation*}
$x_0$, $y_0$, $\nu>2$ are the Merkuriev cut-off parameters,
and 
\begin{widetext}
\begin{equation*}
(h^{\alpha \beta  L}_{l\lambda,l'\lambda'}
\psi_{\beta l' \lambda'}^L) (x_\alpha,y_\alpha)=\int_{-1}^1 dz_\alpha\;
(h^{\alpha \beta  L}_{l\lambda,l'\lambda'} (x_\alpha, y_\alpha, z_\alpha)
\psi_{\beta l' \lambda'}^L (x_\beta,y_\beta)
\end{equation*}
where
\begin{eqnarray*}
h^{\alpha \beta  L}_{l\lambda,l'\lambda'} (x_\alpha, y_\alpha, z_\alpha)
&=&\frac{(-)^{L+\lambda'}}{2} \sum_{k=0}^{(l+\lambda+l'+\lambda')/2} 
(-)^k (2k+1) P_k(z_\alpha)
\sum_{l_1+l_2=l'} \sum_{\lambda_1+\lambda_2=\lambda'} (2l'+1)(2\lambda'+1) 
\\ \nonumber
&& \times \sqrt{(2l+1)(2\lambda+1)} 
\left( \frac{2l'! 2\lambda'!}{2l_1! 2l_2! 2\lambda_1! 2\lambda_2!} \right)^{1/2}
(-)^{\lambda_1} (c_{\beta \alpha})^{l_2+\lambda_1} 
(s_{\beta \alpha})^{l_1+\lambda_2} 
\frac{x_\alpha^{l_2+\lambda_2+1} y_\alpha^{l_1+\lambda_1+1}}
 {x_\beta^{l'+1} y_\beta^{\lambda'+1}}   \\ \nonumber
&& \times \sum_{l'' \lambda''} (2l''+1)(2\lambda''+1) 
\left( \begin{array}{ccc} k & l'' & l \\ 0 & 0 & 0 \end{array} \right) 
\left( \begin{array}{ccc} k & \lambda'' & \lambda \\ 0 & 0 & 0 \end{array} \right) 
\left( \begin{array}{ccc} l'' & \lambda_2 & l_2 \\ 0 & 0 & 0 \end{array} \right) 
\left( \begin{array}{ccc} \lambda'' & \lambda_1 & l_1 \\ 0 & 0 & 0 \end{array} \right) 
\\ \nonumber
&& \times 
\left( \begin{array}{ccc} l & \lambda & L \\ \lambda'' & l'' & k \end{array} \right) 
\left( \begin{array}{ccc} \lambda_1 & \lambda_2 & \lambda' \\
l_1 & l_2 & l' \\
 \lambda'' & l'' & L \end{array} \right).
\end{eqnarray*}
\end{widetext}
The components $\psi_{\alpha l \lambda}(x_\alpha,y_\alpha)$ in (\ref{eqn8}) 
are further expanded in terms of quintic Hermite polynomial splines for each of
the variables $x_\alpha,y_\alpha$.
The wave functions are solved and normalized according to the asymptotic 
wave function 
\cite{hu,kv-hu}
\begin{equation}
\psi^{(\sigma)} \sim f_\sigma + \sum_{\sigma'=1}^{\mbox{open channels}}
\tilde{K}_{\sigma' \sigma} f^{\sigma'}.
\label{eqn9}
\end{equation} 

In the Ore gap, the $f$'s are the product of 
the standard spherical Bessel functions
and the radial part of the bound-state hydrogenic wave functions. 
$\tilde{K}_{\sigma' \sigma}$ differs from the standard $K$-matrix elements
${K}_{\sigma' \sigma}$ by only a kinematic factor.

According to Fig.\ \ref{jacobi} and Eq.\ (\ref{eqn2}) annihilation takes place when
$\vec{x}_2=0$. However, it is clear that the integrals in (\ref{eqn2}) diverge
whenever $\psi_2(\vec{x}_2,\vec{y}_2)$ is involved. Thus, all previous
calculations of $Z_{\mbox{eff}}$ using (\ref{eqn2}) were limited to energies
below the positronium formation threshold. Only recently, Ref.\ \cite{iks}
used an imaginary absorption potential to replace the dynamics of $e^- e^+$
annihilation. In solving the Schr\"odinger equation they found the annihilation
cross section below the $Ps(1s)$ threshold joins smoothly to the positronium
formation cross section just above the threshold. But this imaginary potential
is too week to represent the annihilation dynamics above and away from the
$Ps(1s)$ threshold. According to this model, above the $Ps(1s)$ threshold
only indirect annihilation after formation is possible. Ref.\ \cite{gl}
renormalized the singularity in (\ref{eqn2}) near the $Ps(1s)$ threshold using
the physical $2\gamma$-lifetime of $Ps(1s)$. They also showed the smooth
transition of the annihilation cross section across $Ps(1s)$ threshold. Theirs
is essentially a threshold law, valid only near the threshold. 
Nevertheless, we
will show that the integral 
\begin{equation}
Z_{\mbox{inf}}=\langle \psi_1 | \delta(\vec{r}_2 |\psi_1\rangle
\label{eqn10}
\end{equation} 
is well defined and exists even above the $Ps(1s)$ threshold. In principle, the
divergent part of (\ref{eqn2}) must be renormalized to the physical indirect
annihilation cross section after the positronium is formed. This part is related
to the positronium formation cross sections that we have already reported in
\cite{hu,kv-w-hu}. This report is devoted to the calculation of (\ref{eqn10}).
 To avoid confusion,  we define (\ref{eqn10}) to be
the annihilation coefficient of $e^- e^+$ annihilation in flight.
 Thus (\ref{eqn10})
truly represents the missing open channel,
the neglected dynamics in the quantum mechanical solution (\ref{eqn8}) of the
$e^+ + H$ system or other  $e^+ +$ atom systems.

At $\vec{x}_2=0$, Eq.\ (\ref{eqn6}) gives
\begin{equation*}
\vec{y}_1  =  -{C_{21}}/{S_{21}} \vec{x}_1, \ \ \ \ \ \ 
\vec{y}_2  =  -{1}/{S_{21}} \vec{x}_1,
\end{equation*}
and
\begin{widetext} 
\begin{equation}
Y_{l \lambda}^{L M} (\hat{x}_1, \hat{y}_1)|_{\vec{x}_2=0} =
 \sqrt{\frac{(2l+1)(2\lambda+1)}{4\pi}} 
\left( \begin{array}{ccc} l & \lambda & L \\ 0 & 0 & 0 \end{array} \right)
 Y_{L}^{M} (\hat{y}_2 ), \nonumber
\end{equation}
where $Y_L^M(\hat{y}_2)$ is the spherical harmonic.
Then, 
\begin{eqnarray}
\label{eqn11}
Z_{\mbox{inf}} & =&  \int | \Psi_1(\vec{x}_1, -C_{21}/S_{21} \vec{x}_1)|^2
y_2^2 dy_2 d\Omega_{y_2} \\ \nonumber
 & =& 4\pi \cos^2\delta \sum_{L=0}^\infty (2L+1) \sum_{l \lambda} 
\int dx_1  \frac{1}{|S_{21}|}
\left[ \frac{ 1}{|C_{21}|x_1} \sqrt{\frac{(2l+1)(2\lambda+1)}{4\pi}} 
\left( \begin{array}{ccc} l & \lambda & L \\ 0 & 0 & 0 \end{array} 
\right)  {\Psi_1}_{l \lambda}^L
(x_1,-C_{21}/S_{21}x_1)\right]^2,
\end{eqnarray}
\end{widetext}
where  $\delta$ is the phase shift of the scattering problem
as a result of the normalization according to the asymptotic wave function
(\ref{eqn9}).

When the incoming channel is $e^+ + H$, $\psi_2$ is solely responsible for 
all positronium formation cross section. When $p+Ps$ is the incoming channel,
$\psi_1$ is responsible for all the hydrogen formation cross section . In either
case, annihilation in flight calculated using (\ref{eqn10}) exists.
Thus $Z_{1 \mbox{inf}}$ and $Z_{2 \mbox{inf}}$ are calculated for these two cases
respectively. In Table I we present the $L=0$ contribution to (\ref{eqn11}) for a
number of $k$ values in the Ore gap. Table II gives the corresponding
annihilation cross sections. The point $k=0.71 a_0^{-1}$ is just above the
positronium formation threshold, where the point $k=0.8612 a_0^{-1}$ is near a
Feshbach resonance. Even though the $S$-state annihilation is relatively smooth
in the Ore gap, the sudden increase near a Feshbach resonance is quite
noticeable.

All calculations were carried out with a cut-off radii 
$y_1^{\mbox{max}}=125 a_0$,
$y_2^{\mbox{max}}=228 a_0$, except near the Feshbach resonance where
$y_1^{\mbox{max}}=150 a_0$, $y_2^{\mbox{max}}=250 a_0$ was used.
$Z_{\mbox{inf}}$ is very sensitive to the cut-off radii 
$y_1^{\mbox{max}}$ and $y_2^{\mbox{max}}$ especially when the energies are
close to either the lower or upper thresholds. Converged results are obtained
only for sufficiently large $y_1^{\mbox{max}}$ and $y_2^{\mbox{max}}$
values. This is demonstrated in Table \ref{tab_3} for the case $k=0.71$,
the $Ps(1s)$ formation threshold is located at $k=0.70653$. In contrast, the
scattering cross sections were well converged at a cut-off radii of 
less then $100 a_0$ \cite{hu,kv-w-hu}, 
which were cross-checked by another method
that solves the Faddeev-Merkuriev integral equations by using the
Coulomb-Sturmian separable expansion approach \cite{phhky} and by numerous 
previous calculations using the Schr\"odinger equation.

Calculations with considerable large cut-off radii require substantial 
computer resources, even for relatively low energies such as that used in the
present calculations where all energies are inside the Ore gap between $H(n=2)$
and $Ps(n=1)$ thresholds. The three-body wave functions were obtained by using  
a quintic spline collocation procedure on the Jacobi coordinates
\cite{hu,kv-hu}. The dimension of the converged calculations have reached
$175000$. Calculations at much higher energies and near 
Feshbach resonances are
quite demanding in terms of computer resources. Yet such calculations are
important, for example in the creating of the antihydrogen and the 
studying of its spectroscopy.  In the formation of antihydrogen using the 
process $\bar{p}+Ps \rightarrow \bar{H}+e^-$, $e^- e^+$ annihilation in the
outgoing channel can be an important consideration when excited states are
involved, where numerous Feshbach resonance exist.

Beyond the likely practical applications in antihydrogen research, this
calculation represents the first breakthrough of a long-standing problem in
positron-atom quantum scattering, i.e.\ its inability to account for the 
$e^- e^+$ annihilation above the positronium formation threshold. 
Using Schr\"odinger equation, the calculations in \cite{iks} are valid only in
the immediate vicinity of the positronium formation threshold at
$k=0.70653a_0^{-1}$. The modified Faddeev equations 
enable the explicit separation of the $e^+ + H$
and the $p+Ps$ channels. The annihilation cross sections in the 
$e^+ +H$ channel are
well defined by (\ref{eqn10}). The singularity involving 
the $p+Ps$ channels are renormalized to the indirect annihilation which can be
calculated using
the known physical annihilation cross sections of the positronium themselves and
the respective positronium formation cross sections obtained in the 
solution of (\ref{eqn8}). Thus a quantum mechanical theory which includes the
annihilation channel is obtained.

This work has been supported by the NSF Grant No.Phy-0243740 and by PSC
and SDSC supercomputing centers under grant No.\ MCA96N011P.

\begin{table}
\caption{
$S$-state annihilation parameters in the Ore gap. The wave number $k$'s are
given in $a_0^{-1}$ units, where $a_0$ is the Bohr radius.
\label{tab_1}}
\begin{center}
\begin{tabular}{lllllll}\hline
$k$ & 0.71 & 0.75 & 0.80 & 0.85 & 0.8608 & 0.8612 \\ \hline
$Z_1$ & 0.630 & 0.592 & 0.531 & 0.478 & 0.446 & 0.403 \\ \hline
$Z_2$ & 1.699 & 1.417 & 0.749 & 1.055 & 1.367 & 7.632 \\ \hline
\end{tabular}
\end{center}
\end{table}

\begin{table}
\caption{
$S$-state annihilation cross sections in the Ore gap. 
The $\sigma$'s are
given in $10^{-6}\pi a_0^2$ units.
\label{tab_2}}
\begin{center}
\begin{tabular}{lllllll}\hline
$k$ & 0.71 & 0.75 & 0.80 & 0.85 & 0.8608 & 0.8612 \\ \hline
$\sigma_1$ & 0.3446 & 0.3067 & 0.2578 & 0.2183 & 0.2010 & 0.1819 \\ \hline
$\sigma_2$ & 0.9293 & 0.7337 & 0.3636 & 0.4820 & 0.6167 & 3.442 \\ \hline
\end{tabular}
\end{center}
\end{table}

\begin{table}
\caption{
$S$-state annihilation parameters at $k=0.71$ with $6$ pairs of cut-off radii.
\label{tab_3}}
\begin{center}
\begin{tabular}{lllllll} \hline
$y_1^{\mbox{max}}$ & 50 & 50 & 55 & 50 & 100 & 125 \\ 
$y_2^{\mbox{max}}$ & 120 & 150 & 174 & 180 & 228 & 228 \\ 
$Z_1$ & 0.541 & 0.574 & 0.535 & 0.537 & 0.633 & 0.630 \\ 
$Z_2$ & 0.406 & 101.8 & 2.128 & 0.471 & 1.766 & 1.699 \\ \hline
\end{tabular}
\end{center}
\end{table}

\end{document}